\documentclass[doublecol]{epl2} 
  
\title{\revision{An ubiquitous mechanism for water-like anomalies}}
\shorttitle{} 

\author{Alan Barros de Oliveira\inst{1} \and  Paulo Augusto Netz\inst{2} \and  Marcia C. Barbosa\inst{1}}

\institute{                    
  \inst{1} Universidade Federal do Rio Grande do Sul,
Instituto de F\'{i}sica,
91501-970 -- Porto Alegre, RS -- Brazil\\
  \inst{2} Universidade Federal do Rio Grande do Sul,
Instituto de Qu\'{i}mica,
91501-970 -- Porto Alegre, RS -- Brazil\\
}
\pacs{61.20.Ja}{Computer simulation of liquid structure}
\pacs{61.25.Em}{Molecular liquids}
\pacs{65.20.-w}{Thermal properties of liquids}

\abstract{Using collision driven molecular dynamics  a 
system of spherical particles interacting through an
effective two length scales  potential is studied. The potential can be
tuned by means of a single parameter, $\lambda$, 
from a ramp $\left(\lambda=0.5\right)$  to
a square-shoulder potential $\left(\lambda=1.0\right)$
representing a family of two length scales potential 
in which the shortest interaction distance 
has higher potential energy than the largest interaction
distance. For all the potentials,
ranging between the ramp and the square-shoulder,
density and structural anomalies were found, 
while the diffusion anomaly is found in all but
in the square-shoulder potential. The presence
  anomalies in square-shoulder
potential, not observed in previous simulations, confirm 
the assumption that the two length scales potential 
\revision{is an ubiquitous ingredient for a system to
exhibit water-like anomalies}.}

\begin{document}

\maketitle

\section{Introduction}

Some liquids are known as \emph{anomalous liquids} since they
exhibit unexpected behavior upon 
variations of its thermodynamic conditions. Water is the
canonical example of those anomalous liquids. 

Water expands upon cooling at fixed pressure \cite{Wa64}, 
diffuses faster upon compression at fixed temperature 
\cite{An76,Pr87}, 
and becomes less organized upon increasing density 
-- or equivalently upon compression -- at
constant temperature \cite{Er01}. 
These are the density, diffusion,
and structural anomalies.

The region where these anomalies occur
form nested domes in the density--temperature diagram 
\cite{Er01} -- or pressure-temperature diagram \cite{Ne01}.
The structural anomaly domain 
occupies the outer region of the 
pressure-temperature phase-diagram and the density anomaly region is the 
innermost region. The diffusion anomaly region lies
between these two domains \cite{Er01,Ne01}.
This is the hierarchy of anomalies of water.

Water-like anomalies are also found in 
other liquids. For example, density anomaly
was found experimentally in liquid 
Te \cite{Th76}, S \cite{Sa67,Ke83}, 
and Ge$_{15}$Te$_{85}$ \cite{Ts91}. Simulations
for silica \cite{Sh02}, silicon \cite{Sa03}, and
liquid beryllium \cite{Ag07b}
show that 
density anomaly is also present
in these materials.  
Diffusion  and structural anomalies were found for 
silica \cite{Sh02,An82b,Ts95,Ru06b} and silicon
\cite{Mo05} and structural anomaly is reported
for liquid beryllium \cite{Ag07a,Ag07b}
through simulations. 

In water, the density anomaly is due to the 
hydrogen bonds. The compression of a hydrogen-bonded local environment leads to an increase in entropy, 
or, equivalently, that a local hydrogen-bonded environment possesses a lower density than a non-bonded system would exhibit \cite{URL}.
However, there is no
hydrogen bond in Te, S and Ge$_{15}$Te$_{85}$.

Therefore, how  can the   anomalies  be explained for
bonding and non-bonding systems?
In order to address this question, instead of looking for the
specific mechanism behind the density anomaly in water or in silica, one
has to find the universality behind it.



The simplest framework in which
one can look into the physics of anomalies
is given by the two length scales potentials. The idea is 
that in principle an anisotropic interparticle
potential can be modeled 
as effective potential \cite{HG93}. We will examine  one
class of effective two
length  scales potential  in which 
a compression-based competition arises between
particles population in the second and 
first shells. 

\begin{figure}
\includegraphics[clip=true,scale=0.52]{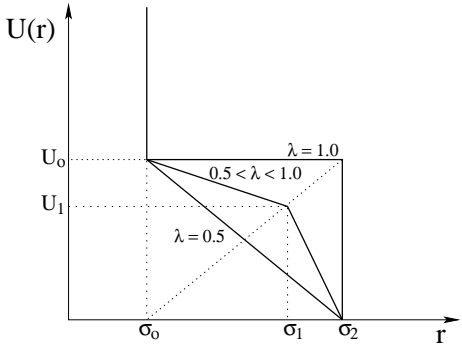}
\caption{Interparticle potential studied in this work.
The potential can be tuned by means of the parameter 
$\lambda$, ranging from a ramp ($\lambda = 0.5$)
to a square-shoulder potential ($\lambda = 1.0$).
See the text for more details.
\label{cap:pot}}
\end{figure}

This assumption was confirmed in a number of isotropic two scales potentials
\cite{Ol06a,Ol06b,Ol07,Mi06b,Wi06,Ca05,Ja99b,Ol08b,Xu06,Ya05,Fo08,
Fr07a} in
which density, diffusion, and structural anomalies were found.
This was also shown to be correct in anisotropic
potentials in which the two length scales emerge from the 
mapping of the anisotropic potential in a equivalent 
spherical symmetric potential \cite{HG93,Ya06}.
Ramp-like potentials have demonstrate to be particularly useful
since they can describe the effective interaction between clusters  
of water molecules \cite{Kr08,Ya08}. In fact, Yan
\emph{at al.} showed a \emph{quantitative} 
agreement between the phase-diagram for the ramp
potential ($\lambda = 0.5$ in Fig. \ref{cap:pot})
and that one for the TIP5P molecular model for water \cite{Ya08}.
The authors show that a central water molecule
interacting with its four nearest neighbors can be modeled effectively
through a ramp potential.

A ramp liquid was also used to mimic water 
in a system composed by a mixture of water and hard-sphere 
particles \cite{Bu07}. 
The solvation thermodynamics of the ramp and hard-sphere 
mixture describes well
the qualitative behavior of the water-like solvation 
thermodynamics \cite{Bu07}.

For studying the water-like anomalies
the exceptional case 
is the square-shoulder potential
in which no anomalies were reported yet (Fig. \ref{cap:pot}
with $\lambda = 1.0$) \cite{Ol08a}. This led to the
idea that anomalies are present in ramp-like potentials (Fig. \ref{cap:pot}
with $\lambda = 0.5$)
but not in shoulder-like potentials.

The aim of this paper is to
propose  that two scales potentials, potentials 
in which two preferred distances 
are present, exhibit water-like anomalies. It will
be shown that in some cases the anomalies
are in  an inaccessible region, as inside
a crystal phase. This is the case for the square-shoulder
potential \cite{Ol08a}.

\section{The Model}

In order to test our assumption
we developed a tunable potential ranging from 
a ramp potential  to a square-shoulder potential.
The potential is given by 
\begin{eqnarray}
U(r) & = & \left\{ \begin{array}{cc}
\infty, & r<\sigma_{o}\\
\phi_{1}(r), & \sigma_{o}<r<\sigma_{1}\\
\phi_{2}(r), & \sigma_{1}<r<\sigma_{2}\\
0, & \sigma_{2}<r,\end{array}\right.
\label{eq:potential}
\end{eqnarray}
\noindent  where 
$\phi_{1}(r)=\left[U_{o}\left(\sigma_{1}-r\right)-U_{1}\left(\sigma_{o}
-r\right)\right]/\left(\sigma_{1}-\sigma_{o}\right)$ and 
$\phi_{2}(r)=U_{1}\left(\sigma_{2}-r\right)\
/\left(\sigma_{2}-\sigma_{1}\right)$.

A single parameter $\lambda$ is used to tune the potential from a ramp
($\lambda=0.5$) to a square-shoulder ($\lambda=1.0$) where 
$\sigma_1 = \sigma_o + \lambda \left( \sigma_2 - \sigma_o \right)$ and
$U_1 = \lambda U_o.$ 

In this work  $\sigma_2/\sigma_o = 1.75$
and the potential was approximated by discrete
steps in the same spirit of Ref. \cite{Ku05},
in such a way that the discrete energy-step
is $\Delta U = 0.025U_o.$  Systems with $0.5\leq\lambda\leq 1.0$ 
were analyzed for density, diffusion, and structural anomalies.

Using the collision driven molecular dynamics 
techniques \cite{Al59}, systems with $N=500$
identical spherical particles of mass $m$, 
interacting through the potential Eq. (\ref{eq:potential})
and $\lambda=0.5,0.6,1.0$ were studied. 
These particles were confined into a cubic box 
with volume $V$ and
periodic boundary conditions. The equilibration
and production times were 500 and 1000 respectively,
in units of $\sigma_o \sqrt{m/U_o}$ (\emph{time units}). The rescaling
of the 
velocities scheme was used for every 2 time units
in order to reach and keep the desired temperature.

Simulations for locating the 
anomalies at the pressure-temperature
phase-diagram were initialized with the system
in the fluid phase. This procedure makes possible to
sample the  metastable
liquid phase inside the  solid phase.

For estimating  the melting line 
the system was initialized with particles in a face centered cubic
configuration. After 500 time units 
it was checked if the system  remains solid 
or if it has melted. This was done by checking the 
diffusion coefficient and the shape of the pair distribution
function.

This 
process gave an estimate of the melting line 
in the pressure-temperature phase-diagram
in excellent agreement
with the one presented in Ref. \cite{Ku05}
for the case in which $\lambda = 0.5$.

\begin{figure}
\includegraphics[clip=true,scale=0.33]{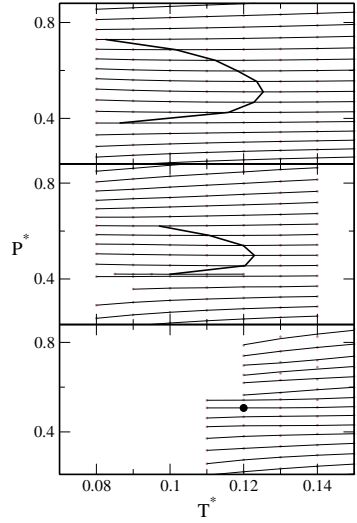}
\caption{Pressure-temperature (P-T) diagram for 
each $\lambda$-case.
Each line in the P-T diagram 
corresponds to an isochore: 
in the upper panel ($\lambda = 0.5$),
from bottom to top, they are 
$\rho^* =$ 0.20, \dots, 0.34, and 0.35.
In the middle ($\lambda = 0.6$)
isochores are $\rho^* =$ 0.19, 0.20,
0.21, 0.22, 0.23, 0.232, 0.24, 
\dots, 0.33, and 0.34.
Finally, in the lower panel ($\lambda = 1.0$)
$\rho^* =$ 0.17, \dots, 0.29, and 0.30.	
The solid bold line connects the 
temperature of maximum density (TMD) points
and it shrinks to a small region 
for $\lambda = 1.0$. See the text for details.
\label{cap:ptall}}
\end{figure}


\begin{figure}
\includegraphics[clip=true,scale=0.33]{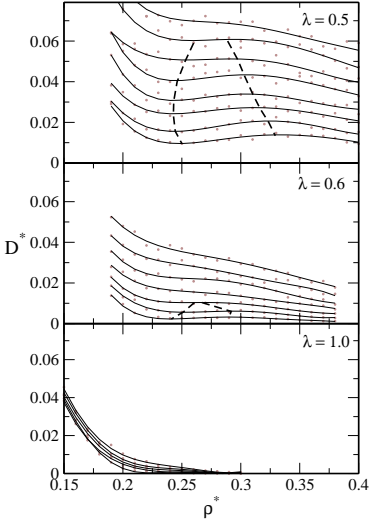}
\caption{Diffusion coefficient against density with
each line corresponding to an isotherm. 
Points are simulated data and continuous lines
are fifth order polynomial fit from the data.
Dashed lines connect maxima and minima
in the D($\rho$, $T =$ constant) functions
and they
have the same meaning as in Fig. \ref{cap:tudojunto}. 
Between these
extrema, diffusion anomalously increases under 
increasing density -- just like water does \cite{An76}.
In the upper panel ($\lambda = 0.5$ case) the isotherms
(from bottom to top)
are $T^* = 0.08$, 0.09, \dots, 0.14, and 0.15.
In the middle panel ($\lambda = 0.6$ case)
the isotherms are $T^* = 0.08$, 0.09, \dots,
0.13, and 0.14. Finally, in the
lower panel ($\lambda = 1.0$ case), 
the isotherms
are $T^* = 0.11$, 0.12,
0.13, 0.14, and 0.15. 
\label{cap:diff}}
\end{figure}

\begin{figure}
\includegraphics[clip=true,scale=0.33]{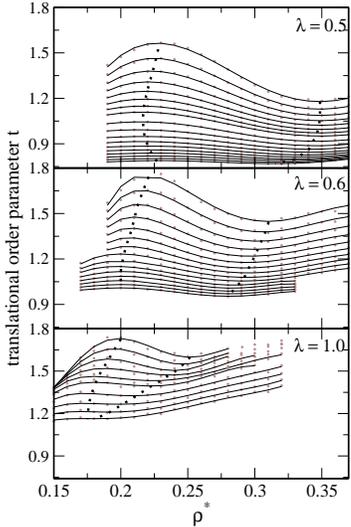}
\caption{Translational order parameter, $t$, versus density for 
fixed temperatures. Points are simulated data
and lines connecting the points are 
fifth order polynomial fit from the data.
Dotted lines bound the region where $t$ decreases under
increasing density and they
have the same meaning as in Fig. \ref{cap:tudojunto}. 
In the upper panel 
($\lambda = 0.5$) we show
(from top to bottom)
the temperatures $T^* = 0.08$, 0.09,
\dots, 0.14, 0.15, 0.17, 0.19, \dots, 0.33,
and 0.35.
In the middle panel ($\lambda = 0.6$) are shown
$T^* = 0.08$, 0.09, \dots, 0.21, and 0.22.
Finally, in the lower panel ($\lambda = 1.0$) temperatures
$T^* = 0.11,$ 0.12, \dots, 0.15, 0.16, 0.18, 0.20, 0.22,
and 0.24 
are shown.
\label{cap:trans}}
\end{figure}


\begin{figure}
\includegraphics[clip=true,scale=0.33]{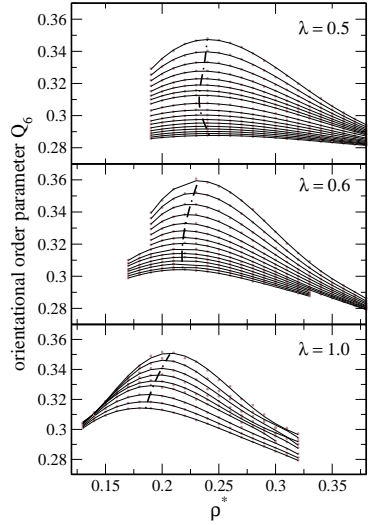}
\caption{Orientational order parameter, $Q_6$, versus density for
fixed temperatures. Points are simulated data
and lines connecting the points are
fifth order polynomial fit from the data.
Dashed-dotted lines connect the maximum values for $Q_6(\rho,T=$ constant)
and they have the same meaning as in Fig. \ref{cap:tudojunto}.
The temperatures shown in the three panels are the same as
in the Fig. \ref{cap:trans}.
\label{cap:Q6}}
\end{figure}


Pressure, $P$, was calculated by
means of virial\cite{Haile}  and diffusion, $D$, 
was derived from the
mean square displacement \cite{Haile}.
The translational order parameter, $t$, was calculated
as \cite{Er01} 
\begin{equation}
t\equiv\int_{0}^{\xi_{c}}|g(\xi)-1|d\xi,
\label{eq:trans}
\end{equation}

\noindent where $\xi\equiv r\rho^{1/3}$ is the interparticle distance $r$
divided by the mean separation between pairs of particles $\rho^{-1/3}. $
$g(\xi)$ is the pair distribution function 
and $\xi_c$ is a cut-off distance.
In this work was used $\xi_c=\rho^{1/3} L/2,$
where $L=V^{1/3}.$  For a completely uncorrelated system (ideal gas) $g=1$
and $t$ vanishes. In a crystal, a translational long-order ($g\ne1$) persists
over long distances making $t$ large.

The orientational order parameter \cite{St83}, $Q_6$, 
was computed using the 
strategy introduced by Yan \emph{el. al} \cite{Ya05}.
$Q_6$ was calculated as follows. First
the spherical harmonics, 
$Y_{lm}^{i}$, associated to each particle $i$ and its $k$ neighbors, 
were obtained by

\begin{equation}
\langle Y_{lm}^{i}\rangle =\frac{1}{k}\sum_{j=1}^{k}Y_{lm}(\theta_{ij},\phi_{ij}).
\label{eq:avYlm}
\end{equation}

The $\langle \dots \rangle$ stands for the average over the $k$ neighbors. Details
on the calculation of $Y_{lm}^{i}$ can be found in \cite{Ol06b,Ya05}.
The orientational order parameter associated to each particle $i$ is
then given by summing  the spherical harmonic over all orders $m$ for a 
fixed degree $\ell$ namely
\begin{equation}
Q_{l}^{i}=\left[\frac{4\pi}{2\ell+1}
\sum_{m=-\ell}^{m=\ell}\left| \langle Y_{lm}^{i}\rangle \right|^{2}\right]^{1/2}.
\label{eq:Qli}
\end{equation}

Then it was chosen $\ell=6$ for characterizing the 
local order \cite{Ya05,Ya06,Ol06b}, so the orientational order
parameter becomes 
\begin{equation}
Q_6=\frac{1}{N}\sum_{i=1}^{N}Q_{6}^{i},
\label{eq:Q6}
\end{equation}
\noindent that is the mean value of $Q_{6}^{i}$ over all particles 
of the system. 

The parameter $Q_6$ assumes its maximum value for a 
perfect crystal and decreases as the system becomes less structured.
For a completely
uncorrelated system (ideal gas) $Q_{6}^{ig}=1/\sqrt{k}.$
In this work $k=12$ neighbors.
For a crystal, the $Q_6$ value depends on
the specific crystalline arrangement and on
the number of neighbors taken into account. 

Pressure,  diffusion, temperature, and density, $\rho = N/V$, 
are given in reduced units as $P^* = P \sigma_{o}^{3}/{U_o}$, 
$D^* = D\left(m/U_o \right)^{1/2}/{\sigma_o}$, 
$T^* =k_B T/U_o$, and  
$\rho^* = \rho \sigma_o^3$. 

\section{The Results}


The potentials  with $0.5\leq \lambda \leq 1.0$ were tested for the presence of: a minimum at the isochores in
the  pressure-temperature phase-diagram; a maximum and 
a minimum in the  diffusion constant as 
a function of  density at constant temperature
diagram,
a maximum and a minimum in the translational order parameter versus density
at constant temperature,
and a maximum in the orientational order parameter against density
at constant temperature. Then the position of the anomalies
were compared with the melting line.

Fig. \ref{cap:ptall} illustrates the pressure-temperature (P-T)
phase-diagram for the potential Eq. (\ref{eq:potential})
with $\lambda = 0.5,$  0.6 and 1.0. Each line corresponds
to an isochore (see the figure caption for details). 
Some  
isochores have minima, which define the 
temperature of maximum density (TMD). For the pressures
and the temperatures inside the TMD, the system 
expands upon cooling under fixed pressure, thus 
this region is known as the density anomaly region \cite{Ku05}.
As $\lambda$ increases, the density anomaly region 
shrinks and reduces to a very small region that we identify with a 
simple point
 the $\lambda = 1.0$ case. This reduction of the
density anomalous region   is consistent with
simulations for   $0.6<\lambda < 1.0$ (not shown)

Fig. \ref{cap:diff} shows the
diffusion constant versus density at fixed temperature
for $\lambda = 0.5$, 0.6 and 1.0.
For the $\lambda<0.7$ cases (not shown)
the diffusion
constant versus density for a fixed temperature
has a local maximum and a local minimum (represented
by a dashed line in Fig. \ref{cap:diff}).
Between these local extrema 
the diffusion coefficient increases
upon increasing density and this region is known as the
diffusion anomaly region.

Figs. \ref{cap:trans} and \ref{cap:Q6}
illustrate the translational and the 
orientational order parameters versus density for
constant temperatures for $\lambda = 0.5,$ 0.6, and 1.0.
The translational order parameter has a local maximum and a
local minimum
for  an interval of temperatures. Between these local extrema,
the parameter $t$ decreases under increasing density. 
An anomalous $t$ parameter 
was observed for  all $\lambda$-cases 
in the stable liquid phase, even  for $\lambda = 1.0$ where
no anomalies were reported before.  
The $t$ anomalous region is bounded 
by the dotted lines
in  Fig. \ref{cap:trans}.

The orientational order parameter,  $Q_6$, 
has a maximum 
in the $Q_6-\rho$ plane for each isotherm as
shown in Fig. \ref{cap:Q6}. 
For densities higher than the density 
of maximum $Q_6$ the orientational
order parameter  decreases as the density increases --
the same trend seen for $t$. 

\begin{figure}
\includegraphics[clip=true,scale=0.38]{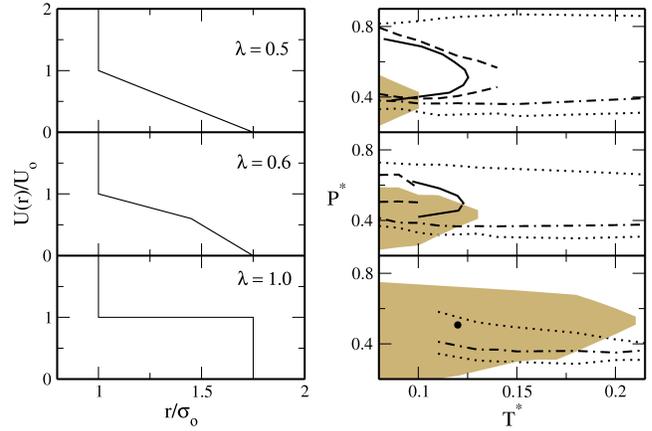}
\caption{Interparticle potential studied in this work
with $\lambda = 0.5, 0.6$ and 1.0 (left) and the
corresponding results we have found (right).
Dotted lines enclose the region where $t$ decreases
upon increasing density and
dashed-dotted line mark the maximum in the $Q_6$  parameter.
The region between the dashed-dotted line and the high pressure
dotted line is known as the structural anomaly region (see
the text for details). Dashed lines bound the region of diffusion anomaly, and
solid line determine the region where density anomaly occurs.
For the case with $\lambda = 1.0$ the density anomaly line shrinks
into a very small region
and the diffusion anomaly region 
lies inside the solid phase, becoming inaccessible.
The shadowed region is bounded by
the estimate of the melting line.
\label{cap:tudojunto}}
\end{figure}

For each temperature the density of maximum $Q_6$ lies
between   
between the density of minimum $t$ and the
density of maximum $t$. 
Consequently for densities between the  density of maximum 
$Q_6$ and the minimum $t$ is 
the structural anomaly region. 
This region is illustrated in Fig. \ref{cap:tudojunto}.


Fig. \ref{cap:tudojunto} summarizes  our findings for  
$\lambda = 0.5,$ 0.6, and 1.0. 
It shows the regions in the P-T 
phase-diagram where the density, diffusion, and structural
anomalies are located. It also show the location of 
the melting line that bounds the region where the system becomes
solid.

For $\lambda = 0.5$ -- the ramp potential case -- the
high pressure dotted line and the dashed-dotted line
bound the region of structural anomalies, where
the translational and orientarional order parameters decrease with density.
A dashed line encloses the diffusion anomaly region  where 
particles diffuse faster upon increasing
density. The 
solid line connects the temperatures of maximum density limiting the
 density anomaly region. 
The border of the shadowed region estimates
the outer limit for the melting line. For the ramp
potential the region of density, diffusion, and
structural anomalies are in the stable region of the pressure
temperature phase-diagram.  All these anomalies
for the ramp case are well
documented \cite{Ku05,Ya06} and a further
discussion on these results is unnecessary.

For the potentials with $\lambda = 0.6$ and 1.0 shown in
Fig. \ref{cap:tudojunto}  the
lines and shadowed region have the same meaning as for
the case with $\lambda = 0.5$. 
Comparing our
results for $\lambda = 0.5,$ 0.6, and 1.0
an interesting trend is observed. Once the potential become 
"harder" -- going from a ramp to a square-shoulder -- three effects are quite evident.

The first effect of the change of $\lambda$ is 
related to the mobility of particles in the system. 
The diffusion anomaly region shrinks as 
$\lambda$ increases, moving to lower temperatures
becoming inaccessible for the case in which $\lambda = 1.0$.
This is in accordance with the results obtained by Netz \emph{et al.} \cite{Ne06}.
These authors have observed that as the discretization of the ramp 
potential becomes less coarser, 
with corresponding increasing in the energy steps, the diffusion 
anomaly region shrinks and migrates to lower temperatures into the density anomaly region. 
Lattice models exhibit this same effect 
since the lattice structure plays
the role of a coarsely discretized energy barriers ambient \cite{Sz07,Gi07}.
In both cases the lines in the 
pressure temperature phase-diagram defining the border between the 
density and diffusion anomalous regions cross for a certain 
choice of parameters what is also observed for $\lambda = 0.6$.

The second effect is related to the 
melting line. As $\lambda$ increases
the estimate of the melting line goes towards high temperatures, engulfing  
the region of water-like anomalies. This could explain why no anomalies
were reported for the case with $\lambda = 1.0$.

The third effect is related to both the 
structural and density anomaly regions. 
As $\lambda$ increases
the temperature of these regions are not drastically affected.  
This is consistent with the results obtained by Netz \emph{et al.} \cite{Ne06},
where the authors
have observed that the discretization of the ramp potential does not
affect the thermodynamics. In the current case,  as  $\lambda$ increases the 
potential not only becomes more discretized but also there is
a change in the potential energy associated with each scale. This 
leads to a shrink in the pressure range of the anomalous region.

\begin{figure}
\includegraphics[clip=true,scale=0.33]{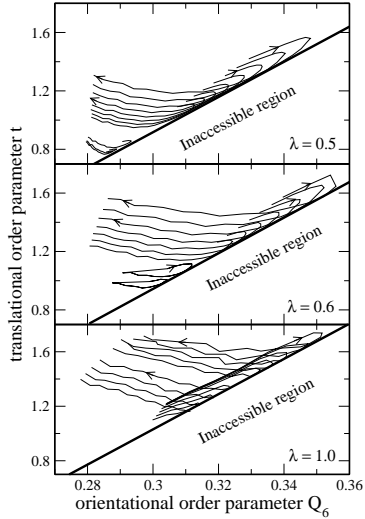}
\caption{Order map for the potential Eq. (\ref{eq:potential})  with
$\lambda = 0.5$, 0.6, and 1.0. 
The arrows indicate
the direction of increasing density for fixed temperatures.
In the upper panel, the temperatures are 
(from top to bottom) $T^{*} = 0.08,$
\dots, 0.14, 0.15, 0.27, 0.33, and 0.35. In the middle
panel, $T^{*} = 0.08,$ \dots, 0.13, 0.14, 0.18, and 0.22.
Finally, in the lower panel, $T^{*} = 0.11,$ \dots,
0.14, 0.15, 0.18, 0.20, and 0.22.
\label{cap:maporder}}
\end{figure}

We can also gain some
insight on the relationship between structure
of the system, shape
of the effective interparticle potential, and presence
of anomalies by analyzing the order map, i.e., the 
$t-Q_6$ plane \cite{Ya05,Ya06,Ol06b,Er01}. 
Fig. \ref{cap:maporder} show our results.

The paths formed
by the points in the order map
for water collapse into a single
line inside the structural anomaly region.
This means that in the water case 
the order parameters are
strongly coupled. For silica \cite{Sh02} 
and beryllium fluoride \cite{Ag07a}, also anomalous 
tetrahedral liquids,
the orientational and translational
order parameters are weakly coupled,
since they develop a two-dimensional
region in the order map of such liquids.
For all $\lambda$ considered in this work we
have a silica- and beryllium fluoride-like
behavior for the paths in the order
map or our model. 
This means that the two scales
potentials are hybrid models, in the sense 
that they can exhibit water-like hierarchy of anomalies
\cite{Ya06,Ol06b} but reproduce
the order map of silica and beryllium fluoride
instead of water.
The understanding of the coupling mechanism
of the order parameters is not clear in the literature,
and we believe this could shade some light into important 
aspects of anomalous fluids.

Next, a remarkable feature of
the order map shown in Fig. \ref{cap:maporder}
is that the inaccessible region is
virtually $\lambda$-independent.
In all panels,
the inaccessible region
is bounded by a straight line given
by $t=a_{\lambda} Q_6 + b_{\lambda}$,
with $a_{0.5} = 12.05$,
$a_{0.6} = 12.20$,
and $a_{1.0} = 12.92$;
$b_{0.5} = -2.70$,
$b_{0.6} = -2.72$,
and $b_{1.0} = -2.84$.
The difference between 
the extreme values of $a_{\lambda}$
and $b_{\lambda}$ quantities
is less than $7.5\%$.
This means that the order in all
$\lambda$-systems respond roughly
constant upon compression. In this sense
we are lead to believe that
the two scales feature,
which is present in all
$\lambda$-cases considered here, is most important
than energetic barriers differences
between them
towards the appearance of anomalies.
Indeed, it was shown that
structure and water-like
anomalies can be linked by means of
the excess entropy \cite{Er06}, stressing
the importance of the role
played by structural parameters
into the understanding of 
the water-like anomalies.


In resume, the three anomalous regions respond differently to the change in
the potential. The diffusion anomalous region shrinks in the
pressure temperature phase-diagram and disappears in $T^*<0.08$ for
$\lambda=0.65$ (not shown); the density anomalous
region shrinks in pressure and reduces to a small region in $\lambda=1.0$;
the structural anomalous region also shrinks in pressure but
it is still in the stable region of the pressure-temperature phase-diagram
for $\lambda=1.0$.
The order map analysis show that the inaccessible
region in the $t-Q_6$ plane is virtually independent
of $\lambda$ and the order parameters 
are uncoupled, differently of water but
similar to silica and beryllium fluoride.

\section{Conclusions}

In this paper we have shown that  two scales 
effective potentials always reproduce
water-like anomalies. In this sense, any liquid material
in which this kind of effective interaction is present 
is able to be an anomalous liquid. In some
cases the anomalous regions are located in the 
pressure-temperature phase-diagram inside the region
where the solid phase is the most stable, being inaccessible either
with experiments or with equilibrium simulations.

In water the two scales distances -- two energetic-competing
preferable distances -- clearly arise
from the formation and breaking of hydrogen bonds. In other
anomalous liquids this process come from different competing interactions 
but the \emph{effective} final 
interparticle potential must be a two scales potential.

We believe that the knowledge of this mechanism can be of
great interest for industries. The domain of this mechanism could lead
to the development of new materials -- polymers, for example --
in which some anomalous properties could be used in the manufacture of
substances close to its supercooled region. 
A polymer in which its molecules diffuses faster upon compression
is an example of a possible application of these findings.

\bibliography{Biblioteca.bib}

\end{document}